\def\wh{wormhole }
\def\beq{\begin{equation}}
\def\eeq{\end{equation}}
\def\bea{\begin{eqnarray}}
\def\eea{\end{eqnarray}}

\def\th{_{_{\rm th}}}

\documentstyle[prl,aps]{revtex}
\begin{document}                                       
\title{Vacuum static Brans-Dicke wormhole}
\author{Luis A. Anchordoqui\thanks{E-mail: doqui@venus.fisica.unlp.edu.ar}, 
A. G. Grunfeld and Diego F. Torres}

\address{Departamento de F\'{\i}sica, Universidad Nacional de La Plata\\
C.C. 67, 1900, La Plata, Argentina\\}

\maketitle

\begin{abstract}

A simple Lorentzian vacuum 
wormhole solution of Brans-Dicke 
gravitation is presented and analysed. 
It is shown that such solution holds for both, 
the Brans-Dicke theory endowed with
torsion (for a value of the coupling parameter 
$\omega > \, 1/2$) and for the vacuum -no torsion- 
case (for $\omega < \,-2$).

\noindent {\it PACS number(s):} 04.20.Gz, 04.50.+h

\end{abstract} 

\vspace{1.2 cm}

Since the renaissance, after the work of Morris and Thorne \cite{motho},
the study of classical \wh solutions have raised an enourmous interest. 
Among the reasons that support this,
one of them is the possibility of constructing
time machines \cite{TM}. 
Another is related to the exoticity of the matter that
threads this kind of geometry, which entails violations of the weak energy
condition (WEC). It is curious that these two motivations also acts as
the main theoretical problems that 
wormhole construction would have. A backward time travel would violate
the chronology protection conjecture while exotic matter, with
negative energy densities, have only appeared at quantum level, with no
macroscopic analogy. In this scheme, studies of possible wormhole solutions
in alternative gravitation was thought of as a way of 
understanding the role of WEC
violation, together with the aim of 
finding phenomena for which different qualitative
behaviors to those of General Relativity may arise. 

In this report we are interested in Brans-Dicke gravitation \cite{BD}. 
This alternative theory, has proved to be a useful tool in the understanding of
early universe models while providing with correct predictions of weak 
field tests and nucleosynthesis.
The case for Lorentzian wormholes in Brans-Dicke theory has been analysed by
Agnese and La Camera \cite{AGNESE} and Nandi et. al. \cite{NANDI}. 
It was shown that three of the four Brans' classes of vacuum solutions 
admit a wormholelike spacetime for convenient choices of their parameters. 
A non-vacuum static solution was also presented where 
WEC was violated by the 
terms corresponding to the scalar, being matter non-exotic 
\cite{TORRES_WH}. 
In addition, 
Euclidean wormholes have also been studied in this gravitational
arena in \cite{EUCLI}.

In what follows, we shall present a static vacuum \wh solution of the 
BD theory endowed with torsion \cite{KIM}. 
This solution, previously
discarded as a general static spherically symmetric 
one due to its unability in
describing the whole range of the radial coordinate \cite{KIMCHO,VB}, 
is now reworked and analysed as a wormhole solution. 
Using afterwards the equivalence of 
the system of field equations between the torsion case and the vacuum one,
we comment on the validity of this same solution for the case of a static
vacuum BD wormhole, by suitably changing the BD coupling $\omega$. 
It is found that this solution is the most simple BD wormhole 
geometry found up to date. Its simplicity and 
well behaved asymptotic properties leads to the possibility of obtaining the 
dependence of the proper wormhole radial coordinate in a fully analytical 
fashion.  

We begin introducing briefly the
skeleton of the BD theory in a spacetime with torsion.
In order to do so, let us first write down 
the action integral for the modified vacuum BD theory,
\begin{equation}
I = \int d{^4}x \sqrt{-g} \left(-\phi R 
+ \omega \frac{ \phi^{,\mu} \phi_{,\mu} }{ \phi} \right),
\label{01}
\end{equation}
where $\phi$ is the BD scalar and $\omega$ is an undetermined constant.
The scalar curvature is that of a Riemann-Cartan spacetime related
to the tetrad field $e_a^\mu$ and the spin conection 
$\omega_{ab}^\mu$ by,
\begin{equation}
R = {e_a}^{\mu} {e_b}^{\nu} {R^{ab}}_{\mu\nu} 
=  {e_a}^{\mu} {e_b}^{\nu} ({\omega^{ab}}_{\mu,\nu} - {\omega^{ab}}_{\nu,\mu}
+ {\omega^a}_{c\nu} {\omega^{cb}}_{\mu} - {\omega^a}_{c\mu} 
{\omega^{cb}}_{\nu}).
\end{equation}       
The torsion field is defined as

\begin{equation}
{F^a}_{\mu\nu} = {e^a}_{\mu,\nu} - {e^a}_{\nu,\mu} +
{\omega^a}_{c\nu} {e^c}_{\mu} - {\omega^a}_{c\mu} {e^c}_{\nu}.
\end{equation}
Henceforth we shall use the conventions of \cite{KIMCHO} unless otherwise
specified. The field equations that follow from such an action are
given as,
\begin{mathletters}
\begin{equation}
R_{\mu\nu} = \omega \phi_{,\mu} \phi_{,\nu}  \phi^{-2} ,
\label{10}
\end{equation}
\begin{equation}
{F^{\mu}}_{\alpha\beta} = \frac{1}{2} \left( \phi_{,\alpha} \delta^{\mu}_{\beta} -
\phi_{,\beta} \delta^{\mu}_{\alpha} \right)  \phi^{-1},
\label{11}
\end{equation}
\begin{equation}
\Delta\phi = {F^{\mu}}_{\lambda\mu} \phi^{,\lambda}.
\label{12}
\end{equation}    
\end{mathletters}
At this stage, and in the sake of completeness, it is important to 
review the basic properties a spacetime needs to obey so as to
display wormholelike features. We begin by introducing the static
spherically symetric line element,
\beq
ds^2 = - e^{2\Phi(r)} dt^2 + e^{2\Lambda(r)} dr^2 + r^2 d\vartheta^2 
+ r^2\,\sin^2\vartheta d\varphi^2
\eeq
where the redshift function $\Phi$ and the shapelike function
$e^{2\Lambda}$ characterize the
wormhole topology and must satisfy,\\
\indent {\it i)} $e^{2\Lambda} \geq 0$ 
throughout the spacetime. This is required
to ensure the finiteness of the proper radial distance defined by $dl
= \,\pm \,e^\Lambda \,dr$. The $\pm$ signs refer to the two
asymptotically flat regions which are connected by the wormhole throat.\\
\indent {\it ii)} The precise 
definition of the wormhole's throat (minimun radius) entails 
a vertical slope of the embedding surface, 
\beq
\lim_{r \rightarrow r\th^+}
\frac{dz}{dr} \, = \lim_{r \rightarrow r\th^+}\, \pm\,
\sqrt{e^{2\Lambda} - 1} = \, \infty.
\label{fo2}
\eeq
\indent {\it iii)} As $l \rightarrow \pm \infty$ 
(or equivalently, $r \rightarrow
\infty$), $e^{2\Lambda} \rightarrow 1$ and $e^{2\Phi} \rightarrow 1$. 
This is the asymptotic
flatness condition on the wormhole spacetime.\\
\indent {\it iv)} $\Phi(r)$ needs to be finite throughout the spacetime to ensure
the absence of event horizons and singularities.\\
\indent {v)} Finally, the  {\em flaring out} condition, that asserts that the
inverse of the embedding function $r(z)$, must satisfy $d^2 r/dz^2 >
0$ at or near the throat. Stated 
mathematically, 
\beq
-\frac{\Lambda^\prime\, e^{-2\Lambda}}{(1-e^{-2\Lambda})^2} > 0.
\label{fo1}
\eeq

We turn back our attention to the previously posed problem. 
Replacing the line element in the field equations, the non-trivial 
set is 

\beq
\left( \Phi^\prime + \frac{\phi^\prime}{2\phi} \right)^\prime +  \left( \Phi^\prime + 
\frac{\phi^\prime}{2\phi} \right) \left( \Phi^\prime - \Lambda^\prime + 
\frac{\phi^\prime}{\phi} + \frac{2}{r} \right) = 0
\label{fe1}
\end{equation}

\beq
(1 + \Phi^\prime - \Lambda^\prime ) 
\left( \Phi^\prime + \frac{\phi^\prime}{2\phi} \right) + (1 -
 \Lambda^\prime) \left( \frac{2}{r} + 
\frac{\phi^\prime}{\phi} \right)   
 - \omega \left( \frac{\phi^\prime}{\phi}\right)^2  = 0
\label{fe2}
\end{equation}

\begin{equation}
\left( \frac{1}{r} + \frac{\phi^\prime}{2\phi} \right) 
\left[ \Phi^\prime - \Lambda^\prime  + \frac{1}{r} \left( \frac{1}{r} 
+ \frac{\phi^\prime}{\phi} \right) 
\right]
+ \left(\frac{1}{r} + \frac{\phi^\prime}{2\phi} \right)^\prime 
- \frac{e^{2 \Lambda}}{r} = 0,
\label{fe3}
\end{equation}

\begin{equation}
\frac{\phi^{\prime\prime}}{\phi^\prime} = (\Lambda^\prime - \Phi^\prime) - \frac{2}{r}.
\label{we}
\end{equation}

As was shown in \cite{KIMCHO}, combining the field equations and 
with a suitable choice of the integration constants, the redshift and 
the shapelike functions could be linked to the BD scalar as follows, 

\beq
\Phi^\prime= - \frac{\phi^\prime}{\phi} 
\hspace{2cm}
e^{2\Lambda} = \alpha \left( \frac{\phi^\prime}{\phi}\right)^2 r^4 
\end{equation} 
with 
$\phi^\prime/\phi = \pm \sqrt{\alpha \,r^2 - \varrho}\,$, being $\alpha$ and 
$\varrho$ positive constants \footnote{Note that $\varrho=(1-2\omega)/4$ 
could be related to the parameter $\beta^2$ of \cite{KIMCHO}.}. 
A trivial integration yields

\begin{equation}
\phi = \phi_0  \exp \left[ \frac{1}{\sqrt{8}} 
{\rm arcsen} \left( - \sqrt{\frac{\varrho}{r^2\,\alpha}} \right) 
\right] 
\end{equation}

\begin{equation}
e^{2\Lambda} = \frac{\alpha r^2}{\alpha r^2 - \varrho}  
\label{lambda}
\end{equation}

\begin{equation}
e^{2\Phi} = {\rm exp} 
\left\{ -2 \left[ \frac{1}{\sqrt{\varrho}}{\rm arcsen}
\left(-{\sqrt\frac{\varrho}{r^2\,\alpha}}\right) 
+ {\cal K} \right] \right\}.
\end{equation}

We shall put into evidence that this solution, valid whenever 
$r \geq \sqrt{\varrho / \alpha}$, and $\omega > \,1/2$, satisfies the
desired properties of traversable wormholes. 
It is easily seen that in the asymptotic limit, the redshift and shapelike
functions tends to unity, being the metric tensor that of a flat Minkowskian
spacetime. Besides, the BD scalar also tends to unity 
if ${\cal K}=-\pi/\sqrt{ \varrho}$. Finally, at the throat, when 
$r \rightarrow  \sqrt{\varrho/\alpha}$, $dz/dr \rightarrow \infty$.

As the $r$ coordinate system has a singularity at the throat, where the
metric coefficient $g_{rr}$ becomes divergent, the spatial
geometry is better studied in terms of the aforementioned 
proper radial coordinate $l$, which can be computed as stated in
property {\it i)}. This yields

\begin{equation}
l = \sqrt{r^2 - \frac{\varrho}{\alpha}}.
\end{equation}
Due to the simple expression for $l(r)$ it is easy to rewrite the
metric tensor in terms of this proper radial distance,

\beq
ds^2=-h(l)\;dt^2 + dl^2 + r^{2}(l) d\Omega ^2,
\eeq
where,

\begin{equation}
h(l) = {\rm exp} 
\left\{ -2 \left[ \frac{1}{\sqrt{\varrho}}{\rm arcsen}
(\sqrt{\frac{\varrho}{\alpha l^2 + \varrho}}) + {\cal K} \right] \right\}.
\end{equation}
and
\begin{equation}
r^{2}(l) = l^2 + \frac{\varrho}{\alpha}.
\end{equation}
Thus, in this well behaved coordinate system, as $l$ increases from $-\infty$
to $0$, $r$ decreases monotonically to a minimum value at the throat;
and as $l$ increases onwards to $+\infty$, $r$ increases monotonically.

As a final remark, we would wish to note that the field equations of
the torsion endowed BD case are equivalent to those of vacuum 
\cite{KIMCHO,VB} by making the substitution 
$\omega \rightarrow -\omega \,\,-3/2$. This
immediately implies that the solution presented is also a
vacuum spherical symmetric solution of Brans-Dicke theory 
without torsion with $\omega < -2$.
In this case it was previously noted that the stress-energy tensor of
the BD field is WEC violating, being the field itself the carrier
of exoticity. That was the case found in the works of Agnese and La Camera
\cite{AGNESE} and Nandi et. al. \cite{NANDI} for others vacuum solutions
and also in \cite{TORRES_WH} for a non-vacuum case.
Thus, wormhole taxonomy is now enriched with a solution that,
being discarded in the past as bad behaved, is now recasted as the most
simple example of wormhole geometry in alternative gravity.

\subsection*{Acknowledgments}

The authors acknowledge S. J. 
Stewart for fruitful discussion and encouragement.
D.F.T. has been partially supported by CONICET and 
L.A.A. by FOMEC.

\end{document}